\pdfoutput=1 

\documentclass[
     a4paper,
     ]{article}

\usepackage{IEK10} 
\usepackage{natbib}

\newcommand{\mytitle}{Graph Neural Networks for Surfactant Multi-Property Prediction}

\newcommand{\affil}{
  \begin{itemize}[leftmargin=3mm, itemsep=0mm]
        \item[$^a$]BASF Personal Care and Nutrition GmbH, Henkelstrasse 67, 40589 Duesseldorf, Germany %
        \item[$^b$]RWTH Aachen University, Process Systems Engineering (AVT.SVT), Aachen, Germany %
		\item[$^c$]Forschungszentrum J\"ulich GmbH, Institute for Energy and Climate Research IEK-10: Energy Systems Engineering, J\"ulich, Germany%
		\item[$^d$]JARA Center for Simulation and Data Science (CSD), Aachen, Germany, Aachen, Germany
  \end{itemize}
}

\def\firstAuthor{Christoforos Brozos}
\newcommand{\myauthor}{
	Christoforos Brozos$^{a,b}$, 
	Jan G. Rittig$^b$,
	Sandip Bhattacharya$^a$,
    Elie Akanny$^a$,
    Christina Kohlmann$^a$,
	Alexander Mitsos$^{d,b,c,*}$ %
}
\newcommand{\correspondingauthor}{
    Corresponding author, E-mail: amitsos@alum.mit.edu
}
\author{\myauthor}

\usepackage[hyphens]{url}
\usepackage[
  colorlinks,
  linkcolor=blue,
  citecolor=blue,
  urlcolor=blue,
  pdftitle={\mytitle},
  pdfauthor={\firstAuthor}
]{hyperref}
\usepackage[capitalise, nameinlink]{cleveref}
\crefname{table}{Tab.}{Tab.}

\begin{document}

	\thispagestyle{firststyle}
	
	\begin{center}
		\begin{large}
			\textbf{\mytitle}
		\end{large} \\
		\vspace{0.1cm}
		\myauthor
	\end{center}
	
	\vspace{-0.4cm}
	
	\begin{footnotesize}
		\affil
	\end{footnotesize}
	
	\vspace{-0.3cm}

	\section*{Abstract}
	
	Surfactants are of high importance in different industrial sectors such as cosmetics, detergents, oil recovery and drug delivery systems. Therefore, many quantitative structure-property relationship (QSPR) models have been developed for surfactants. Each predictive model typically focuses on one surfactant class, mostly nonionics. Graph Neural Networks (GNNs) have exhibited a great predictive performance for property prediction of ionic liquids, polymers and drugs in general. Specifically for surfactants, GNNs can successfully predict critical micelle concentration (CMC), a key surfactant property associated with micellization. A key factor in the predictive ability of QSPR and GNN models is the data available for training. Based on extensive literature search, we create the largest available CMC database with 429 molecules and the first large data collection for surface excess concentration ($\Gamma$$_{m}$), another surfactant property associated with foaming, with 164 molecules. Then, we develop GNN models to predict the CMC and $\Gamma$$_{m}$ and we explore different learning approaches, i.e., single- and multi-task learning, as well as different training strategies, namely ensemble and transfer learning. We find that a multi-task GNN with ensemble learning trained on all $\Gamma$$_{m}$ and CMC data performs best. Finally, we test the ability of our CMC model to generalize on industrial grade pure component surfactants. The GNN yields highly accurate predictions for CMC, showing great potential for future industrial applications.
	
	\vspace{0.1cm}



\section{Introduction}\label{sec:introduction}

\noindent Surfactants are highly relevant molecules used in a wide range of everyday products, such as food, cosmetics, detergents and drugs~\citep{Vieira2021,Shaban2020,Nitschke2007,2005Surf,Adu2020,Szuets2012}. Surface-active agents (surfactants) are amphiphilic molecules with a hydrophobic chain and a polar hydrophilic head. The surfactant molecule orients itself at the interface between two phases with the hydrophobic portion oriented towards the hydrophobic phase (e.g., air/oil) and the hydrophilic portion oriented towards the hydrophilic phase. Due to their structure, surfactants are surface/interface active and they are able to lower the surface/interfacial tension~\citep{Shaban2020,Zhang2019}. Owing to their properties, surfactants are widely used in a variety of applications such as detergents, dispersion stabilizers, foaming agents, lubricants and as pharmaceuticals among many others~\citep{Knop2010,Shaban2020,Schramm2003,Gallou2015}. Surfactants are classified based on their hydrophilic head group, into ionics and nonionics, with the former further classified into anionics, cationics and zwitterionics. Important surfactant properties to various applications are the critical micelle concentration (CMC), the surface tension ($\gamma$), the surface excess concentration ($\Gamma$$_{m}$), the Cloud Point (CP) and the Krafft Temperature (KT)~\citep{Yang2016,AlSabagh2011}. These properties are used to identify new surfactants with desired performance.

Amphiphilic molecules such as surfactants form micelles, i.e., aggregates. Micelles ultimately dictate the surface/interface activity and strongly impacts the solubilization and detergency, or cleaning ability of a surfactant solution~\citep{Patist2001}. The minimum surfactant concentration at which such micelles are formed in a solution is called \emph{critical micelle concentration} (CMC)~\citep{Rosen2012,mukerjee1971critical}. The CMC is an important value in a wide range of applications such as shampoos~\citep{Thompson2023}, bio-materials design for drug delivery systems~\citep{Su2020}, polymeric micelles~\citep{Ghezzi2021,Perumal2022} and oil recovery~\citep{Kumar2019}. In addition, some studies have reported correlation between CMC and surfactant toxicity~\citep{Perinelli2016} and between CMC and foam stability~\citep{Majeed2020}. The CMC is influenced by multiple factors, like temperature, solvent, pH, chemical structure, pressure conditions and size of the tail and head groups~\citep{Katritzky2008,Thiruvengadam2020,Rosen2012}. Determination of CMC is time-consuming and expensive, and several methods can be used like tensiometry~\citep{Dahanayake1986}, refractive index~\citep{mukerjee1971critical}, calorimetry~\citep{Ortona1998}, viscosity and conductivity measurements ~\citep{Jobe1984}. In most methods, a break-point in the measured property (e.g., surface tension or conductivity) vs. surfactant concentration curve is observed, and the CMC is defined to be at that point~\citep{mukerjee1971critical,Perinelli2020,Moulik2021}.

Since surfactants prefer to exist and adsorb at the interfaces, we can define their adsorption effectiveness as the surface excess concentration ($\Gamma$$_{m}$)~\citep{Rosen2012}. $\Gamma$$_{m}$ is an important surfactant property, as it is a measure of surfactant concentration at air/water and oil/water interfaces. Additionally, surface excess concentration has been shown to influence foaming, emulsification and the kinetics of surfactant-induced pore wetting~\citep{Rosen2012,Wang2019}. Like CMC, $\Gamma$$_{m}$  is influenced by surfactant structure and temperature~\citep{MyersAugust2020,Rosen2012}. The implicit calculation of the surface excess concentration is possible, as is the CMC, from a surface tension measurement plot using the Gibbs adsorption equation~\citep{Rosen1982a,Rosen2012}.

Due to the tedious and expensive nature of experiments, the prediction of surfactant properties without experiments has been a focus of research for many years, mainly, through the use of quantitative structure-property relationship (QSPR) models. An overview of QSPR models in surfactants was given by \cite{Hu2010}, with most of the QSPR studies aiming to correlate molecular descriptors with the CMC~\citep{Gaudin2016,Katritzky2008,Mattei2013}. The developed models showed good predictive performance~\citep{Gaudin2016,Katritzky2008,Mattei2013}. Nevertheless, all of them share a common limitation: they are applicable only on a single surfactant class (nonionics, cations etc.). Besides the CMC, similar modeling techniques have been applied to other important surfactant properties like the cloud point of nonionic surfactants or the minimum surface tension at CMC~\citep{Hu2010}. Very recently, the first QSPR model for predicting $\Gamma$$_{m}$ was developed~\citep{Seddon2022}. The QSPR model was trained on data generated from the Szyszkowski equation using experimentally measured SFT-\text{log(c)} profiles and not on the experimental data directly~\citep{Seddon2022}.
However, the authors stress that their QSPR model has limitations in the surfactant categories included, e.g., the model is not applicable to fluorinated surfactants~\citep{Seddon2022}.

In the recent years, graph neural networks (GNNs) have been intensively researched in the field of molecular property prediction, with numerous GNN models existing in the literature, even regarding the same target property~\citep{Yang2019,Wieder2020}. In more detail, GNNs have been applied to a variety of chemical applications such as ferromagnetic materials~\citep{Pasini2022}, the biodegradability of molecules~\citep{rittig2022graph} and the activity coefficients~\citep{SanchezMedina2022,Rittig2022,Felton2021,Rittig2023_GDI}. GNNs are a deep learning technique, where each molecule is represented as a graph, with atoms corresponding to nodes and bonds to edges. In contrast to classical QSPR methods, where molecular descriptors are typically selected manually and therefore require domain knowledge, GNNs can extract, in an automated way, all the necessary structural-related information which are later used in the regression task for property prediction. For surfactants, \cite{Qin2021} used GNNs to predict the CMC of multiple surfactant classes. They showed that GNNs can be efficiently used as an alternative to classical descriptor-based QSPR in surfactants, with very promising results, using a database of 200 molecules, which is relatively small for training a machine learning model.

Herein, we create the largest CMC and $\Gamma$$_{m}$ data sets available and we use them to develop GNN models for their prediction. First, we extend the publicly available CMC data set of \cite{Qin2021} to 429 molecules through an extensive literature search. Then, we construct a second data set of 99 molecules with duplicate values with the aim of investigating possible benefits of transfer learning in CMC prediction. For the $\Gamma$$_{m}$, no publicly available database was found during our research. Therefore, we construct one with 164 surfactant molecules varying from multiple surfactant class types. Compared to the work of \cite{Seddon2022}, we include a wider range of surfactant categories, e.g., fluorinated components, and we consider the impact of counterions on $\Gamma$$_{m}$~\citep{Rosen2012}. Note that the collected data sets include only measurements that can be found in various sources of publicly available literature. 

Furthermore, we establish a GNN model for the prediction of surface excess concentration ($\Gamma$$_{m}$) and a GNN model for the prediction of CMC, both trained on the above mentioned new databases. Additionally, we investigate multi-task learning to overcome data limitations and ensemble learning to enhance the predictive performance. Then, we experimentally measure 3 industrial grade surfactants, previously unseen by the GNN model. Finally, we predict their CMC with our GNN model, which was trained exclusively on literature data with mainly purified surfactants, and demonstrate the model's ability to generalize to unpurified industrial surfactants.

We construct the rest of this work as following: Firstly, we analyze our databases, data sampling procedure and the industrial surfactants used (Section~\ref{sec: main_data}). Thereafter, we give an overview of how GNN models work, the methods we applied and a brief overview of the hyperparameter selection (Section~\ref{sec:gnns_learn_ind}). We then present our results, compare them with previous works and discuss limitations and possible solutions (Section~\ref{sec:res_dis}). Lastly, we summarize our work and suggest possible future improvements (Section~\ref{sec:conclusions}). The test data used for model evaluation is publicly available in our~\href{https://github.com/brozosc/Graph-Neural-Networks-for-Surfactant-Multi-Property-Prediction}{GitHub repository}. The training data set remains property of BASF and could be made available upon request.

\section{Data sets}\label{sec: main_data}						
\noindent 
We now analyze the existing databases and describe our methodology for data collection (Section~\ref{sec:database_analysis}). Following, we discuss the estimation of CMC and how we handled duplicated values for transfer learning (Section~\ref{sec:cmc_evaluation}). Finally, we analyze the collected data sets (Section~\ref{sec:final_data sets}) and we present three industrial grade surfactants for model testing (Section~\ref{sec:industrial_surfactants}).
\subsection{Existing database analysis and data collection}\label{sec:database_analysis} 
\noindent
We started our work by building on the publicly available database of 202 substances from \cite{Qin2021} for CMC prediction. For the surface excess concentration $\Gamma$$_{m}$, we had to exclusively rely on tables in books and publications, such as~\citep{Rosen2012,Dahanayake1986}, because no constructed data set was found in the literature. At first, literature data (CMC and $\Gamma$$_{m}$) was extracted from multiple sources~\citep{Rosen2012,Gaudin2016,mukerjee1971critical} for all the molecules at temperatures between 20-28\textdegree C. Note that since temperature massively impacts both properties, we only focus on the temperature range defined above. We also traced back to the individual articles referenced in the sources mentioned above~\citep{Rosen2012,Gaudin2016,mukerjee1971critical} and extracted additional CMC and $\Gamma$$_{m}$ data. This procedure resulted in an extended data set of 429 distinct substances for CMC. In addition, we simultaneously collected 164 different $\Gamma$$_{m}$ values from multiple sources.
 
\subsection{CMC data collection procedure and duplicate values}\label{sec:cmc_evaluation}
\noindent
During data collection, we often found multiple CMC values for the same surfactant, differing from source to source, due to factors such as purity levels, measuring method of choice and mathematical evaluation of experimental data~\citep{Moulik2021,Perinelli2020}. An example is given in Table~\ref{tab:bromide_example}, where for the same surfactant 4 different values have been reported. The CMC variations are discussed in previous works and remain an issue in surfactant science~\citep{Moulik2021,Perinelli2020,mukerjee1971critical}. 
\begin{table}
    \begin{center}
    	\caption{CMC values of dodecylpyridinium bromide reported in literature at 25\textdegree C.}
        \begin{tabular}{l| c | c }
        \textbf{Method} & \textbf{Value (mM)} & \textbf{Source} \\
        \hline
        Tensiometry & 11.5 & ~\citep{Rosen1982} \\
        Conductometry & 11.3 & ~\citep{Rosen1982} \\
        Conductometry & 10 & ~\citep{Skerjanc1999} \\
        Light scattering & 11.6 & ~\citep{Ford1966}
        \end{tabular}
        \label{tab:bromide_example}
    \end{center}
\end{table}

To handle duplicate values, we decided for a ranking according to the measurement method. Here, we prefer CMC values obtained via tensiometry because one of our targets is to evaluate our model on industrial grade surfactants using CMC values measured through tensiometry. If tensiometry data was not available, we favored data from refractometry measurements since it was found to be reliable by Mukerjee and Mysels~\citep{mukerjee1971critical}. If data only from other methods was available, i.e., neither tensiometry nor refractometry, we also included it into our main data set. All remaining values for a surfactant, i.e., duplicates, are not included in the main data set but rather collected in a separate data set, which we utilize for a transfer learning approach (cf. Section~\ref{sec:transfer_learning_theory}).
We note that for most surfactants where duplicate values exist, the values tend to be very similar to each other and in some cases even equal.

\subsection{Data sets analysis}\label{sec:final_data sets}
\begin{figure}
\centering
\subfloat[CMC]{\includegraphics[height = 6.5cm, width = 8cm]{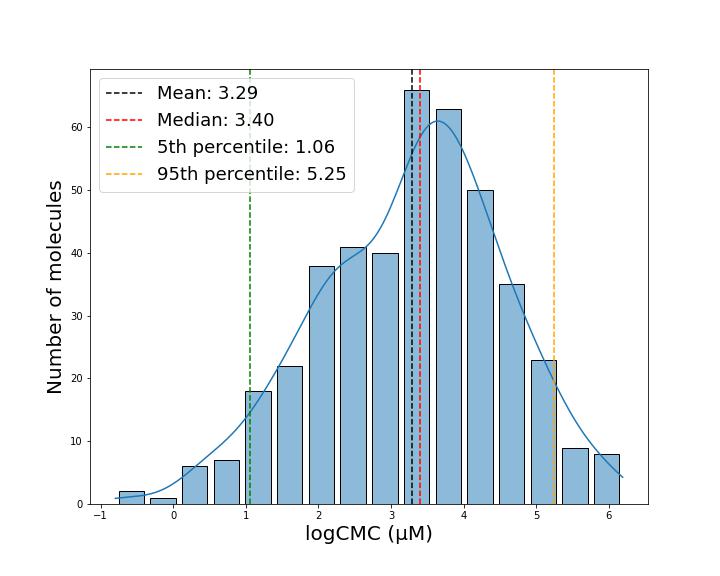}}
\hfill
\subfloat[$\Gamma$$_{m}$]{\includegraphics[height = 6.5cm, width = 8cm]{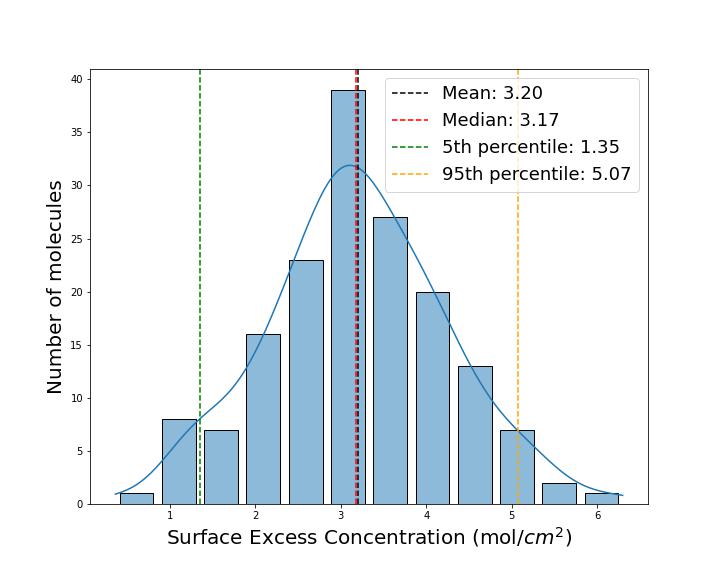}}
\caption{Statistical overview of CMC and $\Gamma$$_{m}$ databases, assembled from literature in this work. Normal distribution of the data set would ensure an even train-validation-test split without artifact.}
\label{fig:statistical_overview}
\end{figure}

\begin{figure}
    \centering
    \includegraphics[height = 8cm, width = 9cm]{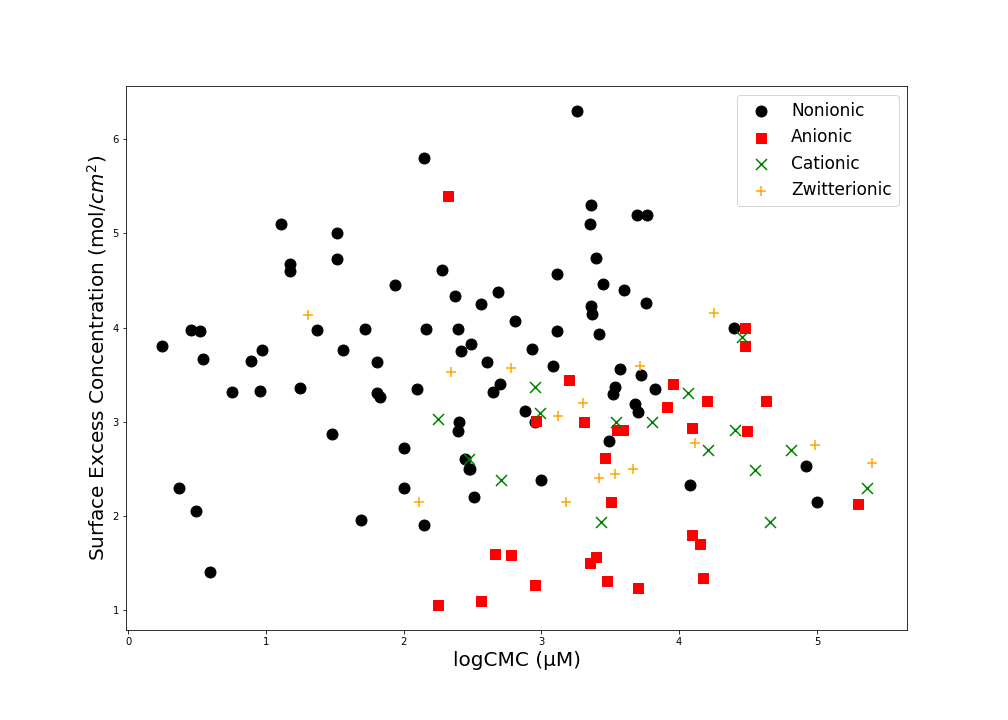}
    \caption{Correlation plot between $\log$ CMC and $\Gamma$$_{m}$ for all surfactant classes. In the plot, 141 surfactants are presented, for which both CMC and $\Gamma$$_{m}$ data was collected from the literature.}
    \label{fig:correlation_plot}
\end{figure}

\noindent
The described sampling process in Section~\ref{sec:cmc_evaluation} led us to construct the three data sets shown in Table~\ref{tab:data sets}, together with a detailed surfactant class distribution. We observe that in the two main databases, nonionic surfactants is the dominant class followed by anionics. This class distribution matches with consumption data of surfactants in 2000~\citep{Rosen2012,MyersAugust2020}, where anionic and nonionic surfactants are the most used in industrialized areas. In other words, the research focus is matching the industrial output. Afterwards, a statistical overview of the target properties, i.e. CMC and $\Gamma$$_{m}$, is presented in Figure~\ref{fig:statistical_overview}. We note that $\Gamma$$_{m}$ shows a natural normal distribution without applying the logarithm. Both data sets have similar mean, median, 5th and 95th percentile values although no comparison between them should be made, as CMC values are scaled. The smallest and biggest values in both data sets are similar too. Finally, a correlation plot between $\log$ CMC and $\Gamma$$_{m}$ is given in Figure~\ref{fig:correlation_plot}, containing surfactants for which both CMC and $\Gamma$$_{m}$ values are collected. 
\begin{table}[b]
	\begin{center}
		\caption{Number of surfactants per class for each database collected in this work. DV = Duplicate Values.}
		\begin{tabular}{l  c  c  c}
		& \textbf{CMC} & \textbf{$\Gamma$$_{m}$} & \textbf{DV-CMC} \\
		\hline 
		Nonionics & 220 & 86 & 19 \\
		
		Anionics & 130 & 44 & 44 \\
		
		Cationics & 55 & 13 & 27 \\
		
		Zwitterionics & 24 & 21 & 9 \\
		\hline
		\textbf{Total substances} & 429 & 164 & 99\\
		\end{tabular}
		\label{tab:data sets}
	\end{center}
\end{table}
\subsection{Industrial Surfactants}\label{sec:industrial_surfactants}
\noindent
With an estimated market size of around \$40 billion in 2020~\citep{FBI2021}, surfactants are also heavily researched in the industry. As Myers~\citep{MyersAugust2020} points out, the majority of academic interest in surfactants, focuses generally on highly purified compounds while the industry is either using complex mixtures to obtain the desired performance or unpurified compounds due to economical reasons. For surface properties like CMC, many authors have noticed the effect of impurities on CMC through the years~\citep{mukerjee1971critical,Moulik2021,Patist2000}. In this study, we examine to what extent GNN models trained exclusively on literature data can generalize for unpurified industrial surfactants. 

Three industrial-grade pure-component surfactants were used, provided by BASF, as obtained from production site without further purification. The main species in these surfactants are (S1) Texapon 842 UP (Sodium Caprylyl Sulfate), (S2) Texapon EHS (Sodium 2-Ethylhexyl Sulfate) and (S3) Texapon K 12 G (Sodium Dodecyl Sulfate). We exclude those three molecules from the training set. According to manufacturing process, we expect the presence of unreacted raw material, alcohols in this case, and reaction by-products.

\section{Methods}\label{sec:gnns_learn_ind}
\noindent
In this section, we first present the fundamentals of a GNN model (Section~\ref{sec:gnn_model}), the general training settings of current works (Section~\ref{sec: train_sett}) and the hyperparameter selection (Section~\ref{sec: hyperparameters}). Afterwards, we refer to the learning techniques applied on this paper (Sections~\ref{sec:single_multi_theory}--\ref{sec:ensemble_learning}). The CMC of the three industrial surfactants was determined by plotting the surface tension as a function of the logarithm of the surfactant concentration. From this plot, two linear regions were determined, which correspond to the linear concentration-dependent and the linear concentration-independent region, respectively. The CMC value is then obtained from the intersection of the straight lines. Finally, for the surface tension measurement a Force Tensiometer – K100 (Krüss, Germany), at 23\textdegree  C was used.

\subsection{Graph Neural Networks}\label{sec:gnn_model}
\begin{figure}
    \centering
    \includegraphics[height = 6.7cm, width = 10 cm]{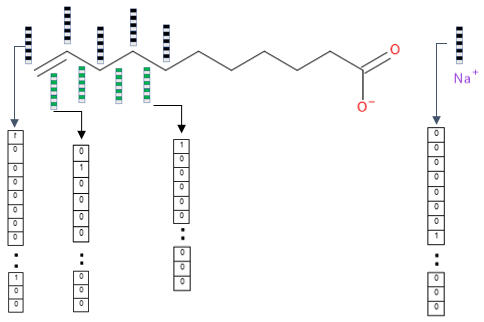}
    \caption{A surfactant molecule represented as undericted graph. In every node (atom) a feature vector (black-white color) of size 30 and in every edge (bond) a feature vector (green-white) of size 12 is assigned. The feature vectors encode chemical information about each individual atom and edge respectively. As can be seen in the atom features vector, different atoms have different entries, which distinguishes them from the other atoms. Similarly, bonds are distinguished through their feaure vector too.}
    \label{fig:mol_to_graph}
\end{figure}

\noindent
In GNN models, every molecule is treated as an undirected graph, where atoms correspond to nodes and bonds to edges. A feature vector, containing chemical information, is assigned to each atom and each edge. Our node and edge features of choice are shown in Table~\ref{tab:node_features} and Table~\ref{tab:edge_features} respectively, motivated from our previous works~\citep{Schweidtmann2020,Rittig2022} and past literature~\citep{Yang2019}. 
Please note that hydrogen atoms are not considered as individual nodes but are implicitly represented in the node feature vector. A surfactant example is given in Figure~\ref{fig:mol_to_graph}.
The molecular graph then passes through graph convolutions, where neighbor information, i.e., neighboring node and edge features, is aggregated for each node in the graph accordingly. The network depth \emph{L}, i.e., the number of graph convolutional layers, defines the neighborhood pool from which structural information will be aggregated. 
We herein use edge-conditioned graph convolutional layers~\citep{Simonovsky2017} and a gated recurrent unit (GRU)~\citep{cho2014learning}, similar to the message passing framework by \cite{gilmer2017neural} and our previous works~\citep{Schweidtmann2020,rittig2022graph,Rittig2022}. 
In contrast to \cite{Qin2021} who used graph convolutional layers only considering node features, we thus explicitly include bond type information in learning the molecular structure which potentially facilitates distinguishing molecules with similar heavy atoms but different bonds, e.g., alkanes versus alkenes. 
After the last graph convolutional layer, the final updated atom feature vectors are pooled into a final molecular fingerprint vector \textbf{h}$_{FP}$~\citep{gilmer2017neural} through a permutation invariant function, i.e., summation of all node vectors. The \textbf{h}$_{FP}$ contains all the necessary structure-related information of a specific molecule required for molecular property prediction, thereby replacing the selected descriptors in classical QSPR techniques mentioned in Section~\ref{sec:introduction}. 

Our model is implemented in the Pytorch Geometric (PyG) framework~\citep{Fey2019}. For the attributed molecular graph generation, we use the SMILES string~\citep{Weininger1988} of each molecule and RDKit (\emph{version 2022.3.5}), an open-source toolkit for cheminformatics.
\subsection{General training settings}\label{sec: train_sett}
\noindent
We use the high-quality data set CMC to define the hyperparameters of our GNN models. For the target property CMC, the $\log$ CMC is calculated and then standardized to a zero mean and a standard deviation of one. The train and test sets are separated randomly in a 85\% - 15\% ratio respectively. For the hyperparameter selection an internal validation set is used, which is a subset of the training set with 20 substances each. The loss function is the mean squared error (MSE) and the optimizer is Adam~\citep{Kingma2014}. In general, we use the same general training settings as in our previous works~\citep{Schweidtmann2020,Rittig2022} and the interested reader can refer to them for more information.
For every modeling approach, which will be introduced in Sections~\ref{sec:single_multi_theory}--\ref{sec:ensemble_learning}, as well as in industrial surfactants application, 40 models were trained on 40 individual training subsets and the results are averaged and reported in Section~\ref{sec:res_dis}.
\begin{table}
	\begin{center}
		\caption{Atom features used in the molecular graph representation. All features are implemented as one-hot-encoding.}
		\begin{tabular}{l | p{7cm} | c}
		\textbf{Feature} & \centering \textbf{Description} & \textbf{Dimension}\\
		\hline
		atom type & \centering atom type (C, N, O, S, F, Cl, Br, Na, I, B, K, H, Li) & 13\\
		is aromatic &\centering if the atom is part of an aromatic system & 1 \\
		hybridization &\centering sp, sp$^2$, sp$^3$, sp$^3$d, or sp$^3$d$^2$ & 5 \\
		\# bonds &\centering number of bonds the atom is involved in & 6 \\
		\# Hs &\centering number of bonded hydrogen atoms & 5 \\
		\hline
		\textbf{Total} & & \textbf{30}\\
		\end{tabular}
		\label{tab:node_features}
	\end{center}
\end{table}

\begin{table}
	\begin{center}
		\caption{Edge features used in the molecular graph representation. All features are implemented as one-hot-encoding.}
		\begin{tabular}{l | c | c}
		\textbf{Feature} & \textbf{Description} & \textbf{Dimension}\\
		\hline
		bond type & single, double, triple, or atomatic  & 4 \\
		is in a ring & whether the bond is part of a ring & 1\\
		conjugated & whether the bond is conjugated & 1 \\
		stereo & none, any, E/Z, or cis/trans & 6 \\
		\hline
		\textbf{Total} & & \textbf{12}\\
		\end{tabular}
		\label{tab:edge_features}
	\end{center}
\end{table}
\subsection{Hyperparameter selection}\label{sec: hyperparameters}
\noindent
With the hyperparameter selection procedure, the aim is to find the suitable hyperparameters of our GNN model. For a robust hyperparameter selection, we test each model in 40 different internal validation sets. We perform a grid search for the following hyperparameters of the GNN model, varying them within the respective ranges: Graph convolutional type  $\in \{NNConv, GINEConv\}$, number of graph convolutional layers $\in \{1,2,3\}$, usage of GRU $\in \{True,False\}$, the batch size $\in \{4,8,16\}$, the initial learning rate  $\in \{0.005, 0.01, 0.05\}$, dimensions of molecular fingerprint and of MLP $\in \{64,128,256\}$. In other words, the hidden layers of the graph convolution part and the hidden layers of the MLP are always the same size. The optimum combination is a GNN architecture with an initial learning rate of 0.005, a hidden state size of 64, a batch size of 16, total graph convolutional layers of 1, the NNConv graph convolutional type and the usage of GRU for the message passing scheme. Our edge feature network, similar to our previous work~\citep{Schweidtmann2020}, consists three layers with the following number of neurons: \#1 12, \#2: 64, and \#3: 4096. The architecture exhibits 306,561 learnable parameters in total. 
\subsection{Single- and Multi-task Learning}\label{sec:single_multi_theory}
\noindent
A surfactant molecule usually has multiple target properties, as we discussed above in Section~\ref{sec:introduction}. The classical learning approach, \emph{single-task learning}, is to train individual models for every property of interest. In single-task learning, model parameters are directly optimized based exclusively on a single target property only, and not transferred to another property prediction task. In that sense, all available QSPR methods for surfactants (discussed in Section~\ref{sec:introduction}) are single-task learning. On the other hand, in \emph{multi-task learning} multiple target properties are simultaneously predicted~\citep{Caruana1997,Zhang2017}. The simultaneous prediction has been shown to improve the modeling accuracy of GNNs in molecular property prediction~\citep{Schweidtmann2020,capela2019multitask,Pasini2022}. Normally during multi-task learning, the graph convolutional layers are shared and individual MLPs are constructed for each target property. The benefits of this approach, are mainly models' ability to generalize, learn faster, reduce overfitting~\citep{ruder2017overview,Crawshaw2020} and data efficiency~\citep{Crawshaw2020}.

In the present work, we investigate the prediction of CMC and of $\Gamma$$_{m}$ with both single- and multi-task learning. Specifically, we develop a single-task learning model for each property individually and multi-task learning models for simultaneous prediction of the properties. Since both of the properties come from the same measurement procedure and therefore are correlated, we expect an improved model accuracy during multi-task learning.

\subsection{Transfer Learning}\label{sec:transfer_learning_theory}
\noindent
Another technique for improving machine learning models is transfer learning~\citep{Hendrycks2019,Zhuang2020}. During transfer learning, a model is usually pre-trained on a data set, for example a synthetic one, and then the model parameters are used to initialize the training on a new unseen data set. This technique is very useful when only small data sets are available. In the field of GNNs, researchers investigated the benefits and limitations of transfer learning~\citep{Grambow2019,Han2021,Kooverjee2022}.\par As we described in Section~\ref{sec:cmc_evaluation}, we collect duplicate values to apply transfer learning to single-task CMC prediction, with the scope of utilizing bigger portions of experimental data from the literature. We use the data set DV-CMC (Table~\ref{tab:data sets}) to pre-train the model, i.e., learn the graph convolutions and MLP parameters, and afterwards we initiate our single-task CMC model with them. All the initialized parameters are optimized based on the CMC data set (Table~\ref{tab:data sets}).
\subsection{Ensemble Learning}\label{sec:ensemble_learning}
\noindent
Training and using single models can lead to under- or/and over-predictions. A well-known technique to mitigate this phenomenon in machine learning is ensemble learning~\citep{Breiman1996,Dietterich2000}. In ensemble learning, multiple models are trained on different subsets of training data set and their final predictions are averaged, resulting in more robust and generalized predictions~\citep{Breiman1996,Dietterich2000,Ganaie2022}.

We use ensemble learning both for our single- and multi-task models mentioned in Section~\ref{sec:single_multi_theory}, by training 40 different models in 40 different subsets of our training data set, in each case. Afterwards, we use the 40 different models to perform predictions in our test set, which are averaged to obtain the final scores.

\section{Results \&  Discussion}\label{sec:res_dis}
\noindent 
In this section, we firstly summarize the predictive performance of our models (Section~\ref{sec: pred_perf}). Afterwards we compare our findings with previous similar work (Section~\ref{sec:prev_works}) and finally we conclude with investigation of model applicability on the selected industrial surfactants (Section~\ref{sec:  ind_pred}). An overview of the performance of the developed models  is reported in Table~\ref{tab:accuracies}. For every task we report the root mean squared error (RMSE), the mean absolute error (MAE) and the variances on the validation and test sets.

\begin{table}
\begin{center}
	\caption{Summary of model accuracy for different predictive tasks over 40 different runs. In each case the standard deviation is also given, except ensemble learning. In the above table we use the following abbreviations: STL = single-task learning, MTL= multi-task learning, TL = transfer learning, EL = ensemble learning, MAE = mean absolute error, RMSE = root mean squared error.}
	\begin{tabular}{l | c  c | c  c  l}
	& \multicolumn{2}{ c }{CMC} &  \multicolumn{2}{ c }{$\Gamma$$_{m}$}\\
	\hline
	& RMSE & MAE & RMSE & MAE \\
	\hline
	STL (val) & 0.27 \textpm{0.027} & 0.2 \textpm{0.026} & 1.02 \textpm{0.169} & 0.85 \textpm{0.15} \\
	STL (test) & 0.33 \textpm{0.033} & 0.25 \textpm{0.026} & 0.8 \textpm{0.143} & 0.57  \textpm{0.146} \\
	STL \& TL (val) & 0.27 \textpm{0.034} & 0.2 \textpm{0.033} \\
	STL \& TL (test) & 0.33 \textpm{0.042} & 0.26 \textpm{0.034}\\
	MTL (val) & 0.26 \textpm{0.075} & 0.2 \textpm{0.053} & 0.3 \textpm{0.121} & 0.26  \textpm{0.116} \\
	MTL (test) & 0.36 \textpm{0.041} & 0.27 \textpm{0.031} & 0.59 \textpm{0.051} & 0.43  \textpm{0.044} \\
	\textbf{STL \& EL (test)} & \textbf{0.28} & \textbf{0.21}  & \textbf{0.76} & \textbf{0.53} \\
	\textbf{MTL \& EL (test)} & \textbf{0.31} & \textbf{0.23} & \textbf{0.56} & \textbf{0.4} \\
	\end{tabular}
	\label{tab:accuracies}
	\end{center}
\end{table}
\subsection{Predictive performance}\label{sec: pred_perf}
\noindent
The single-task GNN model for CMC exhibits an average RMSE of 0.27 on validation set and 0.33 on test set, while the variance is bigger in the test set than in the validation set. For $\Gamma$$_{m}$ the average RMSE in test set is lower than the one in the validation set, with the former equal to 0.85 and the later equal to 1.02. Using the logarithm of $\Gamma$$_{m}$ did not improve the performance. Our model exhibits great performance in predicting the $\log$ CMC, but fails to exhibit similar performance in $\Gamma$$_{m}$ prediction. The reason for the model's under-performance may be the small size of the data set used (140 molecules) for the training and the ambiguous measuring procedures. 
\par In multi-task learning, the GNN model for CMC prediction exhibits an average RMSE of 0.26 on validation set and 0.36 on test set. For $\Gamma$$_{m}$ prediction, the average RMSE in test set is again lower than the one in the validation set, with the former equal to 0.59 and the later equal to 0.43. In the CMC task, the model perform identical with the one in single-task learning, both for validation and test sets, while in the $\Gamma$$_{m}$ task the multi-task model exhibits significantly better performance on the validation set, with the RMSE reducing by 60\%, and improved performance on the test set, with the RMSE reducing by 20\%. Therefore, we conclude that the data limitations of the $\Gamma$$_{m}$ database as single target property can be overcomed by applying multi-task learning. On the other hand, the CMC model did not benefit from the additional data and showed identical results with a slightly higher variance but an overall similar accuracy compared to the single-task learning.
\begin{figure}
\centering
\subfloat[CMC test set: Predicted versus experimental value of log(CMC) in $\mu$M.]{\includegraphics[height = 7cm, width = 7cm]{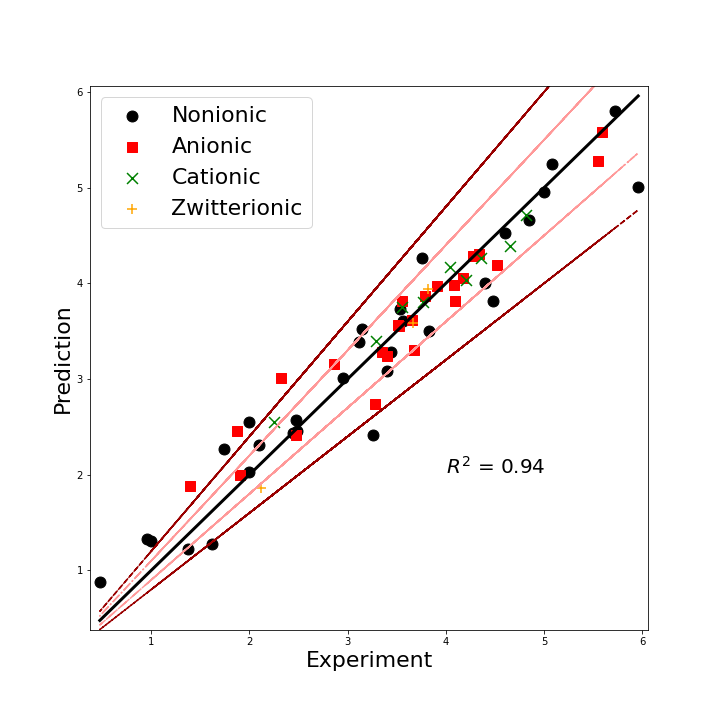}}
\hfill
\subfloat[$\Gamma$$_{m}$ test set Predicted versus experimental value of $\Gamma$$_{m}$ in $mol/cm^2$.]{\includegraphics[height = 7cm, width = 7cm]{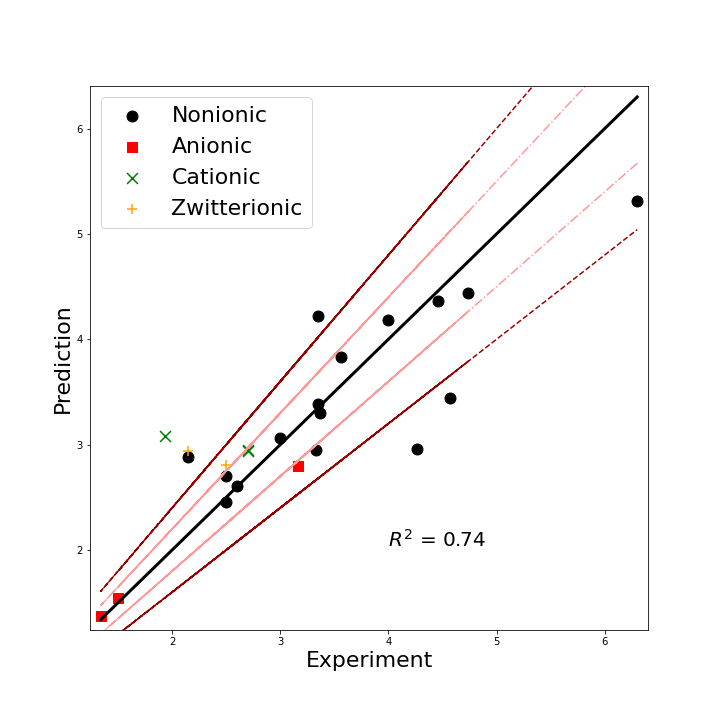}}
\caption{Multi-task GNN ensemble models for (a) CMC and (b) $\Gamma$$_{m}$ for all surfactant classes. The light red dashed lines represent the 10\%  error and the dark red dashed lines the 20\% error.}
\label{fig:parity_ensemble}
\end{figure}

The transfer learning approach, i.e., pre-training the model using the 99 collected duplicate values described in Section~\ref{sec:cmc_evaluation}, is applied only on the single-task CMC model. The RMSE on the validation set remains the same as in single-task learning, equal to 0.27 and on the test set equal to 0.33. Interestingly, the standard deviation increases in both sets. The increase may be due to the broader range of target values for the same property, which leads the model to deviate more from the true value. We observed that transfer learning slightly reduced the final model training time, i.e., the model reached its optimum sooner. Besides the slight reduction of final model training time, using duplicate values for transfer learning led to similar performance.

Ensemble learning, i.e., averaging the predictions of the 40 trained models, slightly reduces the RMSE on test set for single-task learning to 0.28 and for multi-task learning to 0.31 in the CMC case. A similar RMSE reduction is observed on test set for $\Gamma$$_{m}$ accordingly, to 0.76 for single-task learning and to 0.56 for multi-task learning. We use the ensembled results in multi-task CMC and $\Gamma$$_{m}$ learning to draw the parity plots, shown in Figure~\ref{fig:parity_ensemble} on the independent test sets. The parity plots (measured vs predicted values) for CMC and $\Gamma$$_{m}$ show a high determination coefficient for the former, R$^2$$_{CMC}$= 0.94, and moderate one for the latter R$^2$$_{\Gamma_{m}}$= 0.74. We demonstrate that the GNN approach in the present work, can predict CMC and $\Gamma$$_{m}$ across all surfactant classes.

\begin{figure}[b]
	\centering
	\subfloat[]{\includegraphics[height = 1.3cm, width = 3cm]{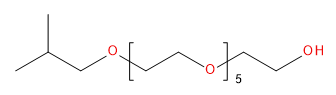}}
	\hfill
	\subfloat[]{\includegraphics[height = 1cm, width = 2.5cm]{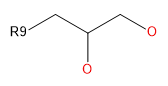}}
	\hfill
	\subfloat[]{\includegraphics[height = 1cm, width = 3cm]{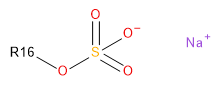}}
	\caption{Outliers in multi-task learning for CMC prediction. Two of them, (a) and (b) belong to nonionic class and the third, (c) to anionic class.}
	\label{fig:cmc_outliers}
\end{figure}

In addition, we present the three components with the highest absolute CMC error in Figure~\ref{fig:cmc_outliers}. For molecule one, the combination of high CMC value and lack of similar molecules, i.e. small alkyl chain with high number of ethylene oxides in the training set, may be the reasons why the model fails to perform well. For molecules two and three, we suspect the measurement may have an impact on the result since identical molecules can be found in the training set. 

\begin{figure}
\centering
\subfloat[]{\includegraphics[height = 2cm, width = 4cm]{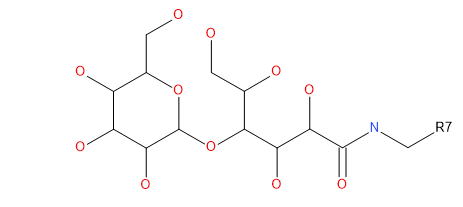}}
\hfill
\subfloat[]{\includegraphics[height = 0.5cm, width = 2.5cm]{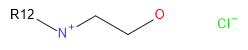}}
\hfill
\subfloat[]{\includegraphics[height = 2cm, width = 4cm]{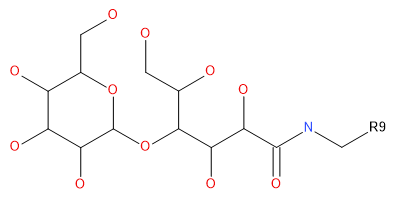}}
\hfill
\subfloat[]{\includegraphics[height = 1cm, width = 2.5cm]{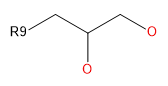}}
\caption{Outliers in multi-task learning for $\Gamma$$_{m}$ prediction. Three of them, (a), (c) and (d) belong to nonionic class and the second, (b) to cationic class.}
\label{fig:gama_outliers}
\end{figure}

Similarly, the four components with the highest absolute $\Gamma$$_{m}$ error are illustrated in Figure~\ref{fig:gama_outliers}. Molecule four is similar as in the CMC case, which supports the measurement impacted hypothesis from before. On the other side molecules one and three have complex structure, where at a similar chemistry is lacking in the training set. Therefore, we can assume that the model fails to capture the property-structure relationship in this case. The same reasoning can be applied to molecule two, where we also lack similar molecules in the training set.

\subsection{Comparison with previous works}\label{sec:prev_works}
\begin{figure}[htb]
\centering
\includegraphics[height = 7cm, width = 10cm]{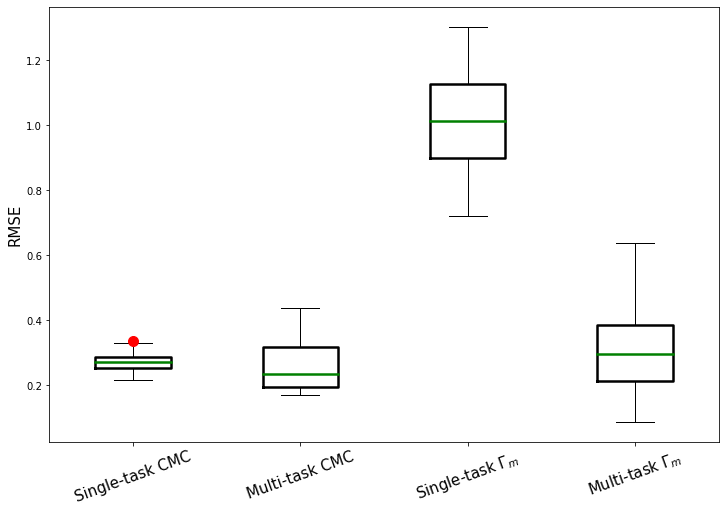}
\caption{Distribution plot of RMSE on internal validation set for all the learning tasks. The boxplots are the results of 40 runs in different internal validation sets. The red points represent the outliers.}
\label{fig:boxplots}
\end{figure}

\noindent
We next compare our results to the work of \cite{Qin2021} and their GNN model for CMC prediction, which used a subset of our CMC training data but is also applicable for a wide choice of surfactant classes. \cite{Qin2021} report a test RMSE of 0.30 which is similar to ours of 0.28. Note that our test set is almost three times the size of the one used by \cite{Qin2021}, so that we cover a higher variance of molecules. Besides the slight RMSE improvement in test set, our model reduces also the average RMSE on the validation set from 0.39 to 0.27. As can be seen in Figure~\ref{fig:boxplots}, no major outliers were observed in the 40 models. Finally, most of the models exhibited a test RMSE in the range of 0.26-0.29 while \cite{Qin2021} reported a broader test RMSE range of 0.28-0.45.

Overall, the comparison shows that the general performance of our model on the CMC is similarly high with previous ones, although a direct quantitative comparison is not possible due to the different data sets used. As there is no model for predicting surface excess concentration directly from experimental data (cf. Section~\ref{sec:introduction}), we are unable to compare our results for $\Gamma$$_{m}$ with other works.
    
\subsection{Industrial Surfactants}\label{sec: ind_pred}
\noindent
We then apply our developed GNN model to predict on the three pure component industrial surfactants described in Section~\ref{sec:industrial_surfactants}. According to the above discussed results, the best learning approach for CMC is the combination of single-task with ensemble learning. We use the 40 trained models from Section~\ref{sec: pred_perf} to perform ensembled predictions on the three surfactants. The predicted log CMC values, as well the experimental measured ones, are given in Table~\ref{tab:industrial_surfactants} for comparison. For all of the three, the predicted values are very close to the measured ones. Overall, the data indicates that the developed GNN model trained on literature data can accurately predict the CMCs for all three single-molecule industrial unpurified surfactants. 
\par We note that similar molecules are used in the training set and none of the three exhibits high structural complexity. On the other hand, the impurities effects on CMC are not learnt from the model and their implementation could potentially be future area of research. Future work could also focus on testing the applicability of GNN model in research of bio-surfactants, with many of them naturally exhibit high structural complexity.

\begin{table}[t!]
	\begin{center}
		\caption{Comparison of predicted values from our single-task GNN with ensemble versus experimentally calculated values for three selected industrial grade surfactants. For the predicted values, the standard deviation over 40 individual runs is also given. The above values, are the logarithmic ones.}
	\begin{tabular}{l | c |  c}
	 &  \textbf{Predicted CMC ($\mu$M)}  & \textbf{Measured CMC ($\mu$M)}\\
	\hline
	S1 & 4.87 \textpm{0.11} & 4.88 \\
	\hline
	S2 & 4.95 \textpm{0.17} & 4.98 \\
	\hline
	S3 & 3.91 \textpm{0.08} & 3.86 \\
	\end{tabular}
	\label{tab:industrial_surfactants}
	\end{center}
\end{table}

\section{Conclusions and future work}\label{sec:conclusions}
\noindent 
We apply GNNs to pure component surfactants to predict CMC and $\Gamma$$_{m}$. Based on extensive literature scanning, we generate a database for CMC with double the size of existing. We also construct a data set for $\Gamma$$_{m}$. As GNNs have been successfully used for CMC prediction~\citep{Qin2021}, we extend the GNN architecture to simultaneously predict the surface excess concentration $\Gamma$$_{m}$ in a multi-task learning, thereby utilizing correlations between these two properties. Furthermore, we collect additional CMC values from the literature and investigate if transfer learning can increase the model accuracy. 

All GNN models exhibit high-accuracy CMC predictions, on a comparable level to a recently developed GNN model by \cite{Qin2021} but for an extended spectrum of surfactants. 
For $\Gamma$$_{m}$ on the other hand, the single-task GNNs fail to capture the property-structure relationship; here, we find that multi-task GNNs effectively utilizes the CMC data to substantially enhance prediction accuracy for $\Gamma$$_{m}$. 
In all cases, ensemble learning increases the prediction accuracy.
For transfer learning, however, we observe no improvements in the model accuracy. 
Finally, we test the best GNN model for CMC on three unpurified industrial surfactants and find highly accurate predictions matching our laboratory measurements, thereby indicating strong potential for further industrial applications.

Furthermore, our GNN models are subject to certain limitations. As is typically the case for ML models, the applicability of our GNN models is limited to surfactants with similar structure as the ones contained in the training data set. Stereochemistry is also not taken into account in this work, for example in the case of n-dodecyl-D-maltoside we only use the CMC of the $\alpha$ isomer~\citep{Pagliano2012}. Another limitation, shortly mentioned above, is not including information regarding the purity of each compound. We only considered highly purified compounds reported in literature and future research should focus in incorporating surface active impurities to the GNN model. 
\par Future work could extend the relative small database for surface excess concentration $\Gamma$$_{m}$, thus yielding higher performance predictive models. For $\Gamma$$_{m}$ noise is often encountered in reported values due to it is implicit calculation through various approaches of the Gibbs adsorption equation~\citep{Rosen2012}. This noise prohibits the GNN model to better capture the structure-property relationship. Finally, prediction of further surfactant properties based on the structure would be highly interesting for surfactant formulators.

\section*{Acknowledgments}

\noindent The BASF authors (C. Brozos, S. Bhattacharya, E. Akanny and C. Kohlmann) were funded by the BASF Personal Care and Nutrition GmbH.
J. G. Rittig and A. Mitsos acknowledge funding from the Deutsche Forschungsgemeinschaft (DFG, German Research Foundation) – 466417970 – within the Priority Programme ``SPP 2331: Machine Learning in Chemical Engineering''.
Additionally, J. G. Rittig acknowledges the support of the Helmholtz School for Data Science in Life, Earth and Energy (HDS-LEE).

\section*{Data availability}
\noindent All python scripts and the test data used in this work are available as open-source at~\url{https://github.com/brozosc/Graph-Neural-Networks-for-Surfactant-Multi-Property-Prediction}.

\section*{Author contribution}
\noindent
\textbf{Christoforos Brozos}: Conceptualization, Methodology, Software, Data curation, Validation, Formal analysis, Writing - Original Draft, Writing - Review \& Editing, Visualization

\noindent\textbf{Jan G. Rittig}: Conceptualization, Methodology, Software, Formal analysis, Writing - Review \& Editing

\noindent\textbf{Sandip Bhattacharya}: Conceptualization, Methodology, Formal analysis, Supervision, Writing - Review \& Editing

\noindent\textbf{Elie Akanny}: Conceptualization, Methodology, Experimental methodology \& measurements, Writing - Review \& Editing

\noindent\textbf{Christina Kohlmann}: Writing - Review \& Editing, Supervision, Funding acquisition

\noindent\textbf{Alexander Mitsos}: Writing - Review \& Editing, Supervision, Funding acquisition

  \clearpage

  \bibliographystyle{apalike}
  \renewcommand{\refname}{Bibliography}
  \bibliography{literature.bib}

\begin{thebibliography}{}

\bibitem[Adu et~al., 2020]{Adu2020}
Adu, S.~A., Naughton, P.~J., Marchant, R., and Banat, I.~M. (2020).
\newblock Microbial biosurfactants in cosmetic and personal skincare
  pharmaceutical formulations.
\newblock {\em Pharmaceutics}, 12.

\bibitem[Al-Sabagh et~al., 2011]{AlSabagh2011}
Al-Sabagh, A., Nasser, N., Migahed, M., and Kandil, N. (2011).
\newblock Effect of chemical structure on the cloud point of some new non-ionic
  surfactants based on bisphenol in relation to their surface active
  properties.
\newblock {\em Egyptian Journal of Petroleum}, 20(2):59--66.

\bibitem[Breiman, 1996]{Breiman1996}
Breiman, L. (1996).
\newblock Bagging predictors.
\newblock {\em Machine Learning}, 24(2):123--140.

\bibitem[Capela et~al., 2019]{capela2019multitask}
Capela, F., Nouchi, V., Deursen, R.~V., Tetko, I.~V., and Godin, G. (2019).
\newblock Multitask learning on graph neural networks applied to molecular
  property predictions.
\newblock arXiv preprint arXiv:1910.13124.

\bibitem[Caruana, 1997]{Caruana1997}
Caruana, R. (1997).
\newblock Multitask learning.
\newblock {\em Machine Learning}, 28(1):41--75.

\bibitem[Cho et~al., 2014]{cho2014learning}
Cho, K., van Merrienboer, B., Gulcehre, C., Bahdanau, D., Bougares, F.,
  Schwenk, H., and Bengio, Y. (2014).
\newblock Learning phrase representations using rnn encoder-decoder for
  statistical machine translation.
\newblock arXiv preprint arXiv:1406.1078.

\bibitem[Crawshaw, 2020]{Crawshaw2020}
Crawshaw, M. (2020).
\newblock Multi-task learning with deep neural networks: A survey.
\newblock arXiv preprint arXiv:2009.09796.

\bibitem[Dahanayake et~al., 1986]{Dahanayake1986}
Dahanayake, M., Cohen, A.~W., and Rosen, M.~J. (1986).
\newblock Relationship of structure to properties of surfactants. 13. surface
  and thermodynamic properties of some oxyethylenated sulfates and sulfonates.
\newblock {\em J. Phys. Chem.}, 90(11):2413--2418.

\bibitem[Dietterich, 2000]{Dietterich2000}
Dietterich, T.~G. (2000).
\newblock Ensemble methods in machine learning.
\newblock In {\em Proceedings of the First International Workshop on Multiple
  Classifier Systems}, MCS '00, page 1–15, Berlin, Heidelberg.
  Springer-Verlag.

\bibitem[Felton et~al., 2022]{Felton2021}
Felton, K.~C., Ben-Safar, H., and Alexei, A.~A. (2022).
\newblock Deepgamma: A deep learning model for activity coefficient prediction.
\newblock {\em 1st Annual AAAI Workshop on AI to Accelerate Science and
  Engineering (AI2ASE)}.

\bibitem[Fey and Lenssen, 2019]{Fey2019}
Fey, M. and Lenssen, J.~E. (2019).
\newblock Fast graph representation learning with pytorch geometric.
\newblock arXiv preprint arXiv:1903.02428v3.

\bibitem[Ford et~al., 1966]{Ford1966}
Ford, W., Ottewill, R., and Parreira, H. (1966).
\newblock Light-scattering studies on dodecylpyridinium halides.
\newblock {\em Journal of Colloid and Interface Science}, 21(5):522--533.

\bibitem[{Fortune Business Insights}, 2021]{FBI2021}
{Fortune Business Insights} (2021).
\newblock {Surfactants Market Size, Share \& COVID-19 Impact Analysis, By Type
  (Anionic, Nonionic, Cationic, and Amphoteric), By Application (Home Care,
  Personal Care, Textile, Food \& Beverages, Industrial \& Institutional
  Cleaning, Plastics, and Others), and Regional Forecast, 2021-2028}.
\newblock
  \url{https://www.fortunebusinessinsights.com/surfactants-market-102385}
  (accessed 30-12-2023).

\bibitem[Gallou et~al., 2015]{Gallou2015}
Gallou, F., Isley, N., Ganic, A., Onken, U., and Parmentier, M. (2015).
\newblock Surfactant technology applied toward an active pharmaceutical
  ingredient: more than a simple green chemistry advance.
\newblock {\em Green Chem.}, 18.

\bibitem[Ganaie et~al., 2022]{Ganaie2022}
Ganaie, M., Hu, M., Malik, A., Tanveer, M., and Suganthan, P. (2022).
\newblock Ensemble deep learning: A review.
\newblock {\em Engineering Applications of Artificial Intelligence},
  115:105151.

\bibitem[Gaudin et~al., 2016]{Gaudin2016}
Gaudin, T., Rotureau, P., Pezron, I., and Fayet, G. (2016).
\newblock New qspr models to predict the critical micelle concentration of
  sugar-based surfactants.
\newblock {\em Ind. Eng. Chem. Res.}, 55(45):11716--11726.

\bibitem[Ghezzi et~al., 2021]{Ghezzi2021}
Ghezzi, M., Pescina, S., Padula, C., Santi, P., {Del Favero}, E., Cantù, L.,
  and Nicoli, S. (2021).
\newblock Polymeric micelles in drug delivery: An insight of the techniques for
  their characterization and assessment in biorelevant conditions.
\newblock {\em Journal of Controlled Release}, 332:312--336.

\bibitem[Gilmer et~al., 2017]{gilmer2017neural}
Gilmer, J., Schoenholz, S.~S., Riley, P.~F., Vinyals, O., and Dahl, G.~E.
  (2017).
\newblock Neural message passing for quantum chemistry.
\newblock {\em 34th International Conference on Machine Learning, ICML 2017},
  3:2053--2070.

\bibitem[Grambow et~al., 2019]{Grambow2019}
Grambow, C., Li, Y.-P., and Green, W.~H. (2019).
\newblock Accurate thermochemistry with small data sets: A bond additivity
  correction and transfer learning approach.
\newblock {\em J. Phys. Chem. A}, 123(27):5826--5835.

\bibitem[Han et~al., 2021]{Han2021}
Han, X., Huang, Z., An, B., and Bai, J. (2021).
\newblock Adaptive transfer learning on graph neural networks.
\newblock In {\em Proceedings of the 27th ACM SIGKDD Conference on Knowledge
  Discovery \& Data Mining}, KDD '21, page 565–574, New York, NY, USA.
  Association for Computing Machinery.

\bibitem[Hendrycks et~al., 2019]{Hendrycks2019}
Hendrycks, D., Lee, K., and Mazeika, M. (2019).
\newblock Using pre-training can improve model robustness and uncertainty.
\newblock In Chaudhuri, K. and Salakhutdinov, R., editors, {\em Proceedings of
  the 36th International Conference on Machine Learning}, volume~97 of {\em
  Proceedings of Machine Learning Research}, pages 2712--2721. PMLR.

\bibitem[Hu et~al., 2010]{Hu2010}
Hu, J., Zhang, X., and Wang, Z. (2010).
\newblock A review on progress in qspr studies for surfactants.
\newblock {\em International journal of molecular sciences}, 11:1020--47.

\bibitem[Jobe and Reinsborough, 1984]{Jobe1984}
Jobe, D.~J. and Reinsborough, V.~C. (1984).
\newblock Micellar properties of sodium alkyl sulfoacetates and sodium dialkyl
  sulfosuccinates in water.
\newblock {\em Canadian Journal of Chemistry}, 62:280--284.

\bibitem[Katritzky et~al., 2008]{Katritzky2008}
Katritzky, A.~R., Pacureanu, L.~M., Slavov, S.~H., Dobchev, D.~A., and
  Karelson, M. (2008).
\newblock Qspr study of critical micelle concentrations of nonionic
  surfactants.
\newblock {\em Ind. Eng. Chem. Res.}, 47(23):9687--9695.

\bibitem[Škerjanc et~al., 1999]{Skerjanc1999}
Škerjanc, J., Kogej, K., and Cerar, J. (1999).
\newblock Equilibrium and transport properties of alkylpyridinium bromides.
\newblock {\em Langmuir}, 15(15):5023--5028.

\bibitem[Kingma and Ba, 2014]{Kingma2014}
Kingma, D.~P. and Ba, J. (2014).
\newblock Adam: A method for stochastic optimization.
\newblock arXiv preprint arXiv:1412.6980.

\bibitem[Knop et~al., 2010]{Knop2010}
Knop, K., Hoogenboom, R., Fischer, D., and Schubert, U. (2010).
\newblock Poly(ethylene glycol) in drug delivery: Pros and cons as well as
  potential alternatives.
\newblock {\em Angewandte Chemie International Edition}, 49(36):6288--6308.

\bibitem[Kooverjee et~al., 2022]{Kooverjee2022}
Kooverjee, N., James, S., and van Zyl, T. (2022).
\newblock Investigating transfer learning in graph neural networks.
\newblock {\em Electronics}, 11(8).

\bibitem[Kumar and Mandal, 2019]{Kumar2019}
Kumar, A. and Mandal, A. (2019).
\newblock Critical investigation of zwitterionic surfactant for enhanced oil
  recovery from both sandstone and carbonate reservoirs: Adsorption,
  wettability alteration and imbibition studies.
\newblock {\em Chemical Engineering Science}, 209:115222.

\bibitem[Majeed et~al., 2020]{Majeed2020}
Majeed, T., S{\o}lling, T.~I., and Kamal, M.~S. (2020).
\newblock Foamstability: The interplay between salt-, surfactant- and critical
  micelle concentration.
\newblock {\em Journal of Petroleum Science and Engineering}, 187:106871.

\bibitem[Mattei et~al., 2013]{Mattei2013}
Mattei, M., Kontogeorgis, G.~M., and Gani, R. (2013).
\newblock Modeling of the critical micelle concentration (cmc) of nonionic
  surfactants with an extended group-contribution method.
\newblock {\em Ind. Eng. Chem. Res.}, 52(34):12236--12246.

\bibitem[Moulik et~al., 2021]{Moulik2021}
Moulik, S.~P., Rakshit, A.~K., and Naskar, B. (2021).
\newblock Evaluation of non-ambiguous critical micelle concentration of
  surfactants in relation to solution behaviors of pure and mixed surfactant
  systems: A physicochemical documentary and analysis.
\newblock {\em Journal of Surfactants and Detergents}, 24(4):535--549.

\bibitem[Mukerjee and Mysels, 1971]{mukerjee1971critical}
Mukerjee, P. and Mysels, K.~J. (1971).
\newblock Critical micelle concentrations of aqueous surfactant systems.
\newblock Technical report, National Standard reference data system.

\bibitem[Myers, 2020]{MyersAugust2020}
Myers, D. (August 2020).
\newblock {\em Surfactant Science and Technology, 4th Edition}.
\newblock John Wiley \& Sons, Ltd.

\bibitem[Nitschke and Costa, 2007]{Nitschke2007}
Nitschke, M. and Costa, S. (2007).
\newblock Biosurfactants in food industry.
\newblock {\em Trends in Food Science \& Technology}, 18(5):252--259.

\bibitem[Ortona et~al., 1998]{Ortona1998}
Ortona, O., Vitagliano, V., Paduano, L., and Costantino, L. (1998).
\newblock Microcalorimetric study of some short-chain nonionic surfactants.
\newblock {\em Journal of Colloid and Interface Science}, 203(2):477--484.

\bibitem[Pagliano et~al., 2012]{Pagliano2012}
Pagliano, C., Barera, S., Chimirri, F., Saracco, G., and Barber, J. (2012).
\newblock Comparison of the $\alpha$ and $\beta$ isomeric forms of the
  detergent n-dodecyl-d-maltoside for solubilizing photosynthetic complexes
  from pea thylakoid membranes.
\newblock {\em Biochimica et Biophysica Acta (BBA) - Bioenergetics},
  1817(8):1506--1515.
\newblock Photosynthesis Research for Sustainability: From Natural to
  Artificial.

\bibitem[Pasini et~al., 2022]{Pasini2022}
Pasini, M.~L., Zhang, P., Reeve, S.~T., and Choi, J.~Y. (2022).
\newblock Multi-task graph neural networks for simultaneous prediction of
  global and atomic properties in ferromagnetic systems*.
\newblock {\em Machine Learning: Science and Technology}, 3(2):025007.

\bibitem[Patist et~al., 2000]{Patist2000}
Patist, A., Bhagwat, S.~S., Penfield, K.~W., Aikens, P., and Shah, D.~O.
  (2000).
\newblock On the measurement of critical micelle concentrations of pure and
  technical-grade nonionic surfactants.
\newblock {\em Journal of Surfactants and Detergents}, 3(1):53--58.

\bibitem[Patist et~al., 2001]{Patist2001}
Patist, A., Oh, S., Leung, R., and Shah, D. (2001).
\newblock Kinetics of micellization: Its significance to technological
  processes.
\newblock {\em Colloids and Surfaces A: Physicochemical and Engineering
  Aspects}, 176:3--16.

\bibitem[Perinelli et~al., 2016]{Perinelli2016}
Perinelli, D.~R., Cespi, M., Casettari, L., Vllasaliu, D., Cangiotti, M.,
  Ottaviani, M.~F., Giorgioni, G., Bonacucina, G., and Palmieri, G.~F. (2016).
\newblock Correlation among chemical structure, surface properties and
  cytotoxicity of n-acyl alanine and serine surfactants.
\newblock {\em European Journal of Pharmaceutics and Biopharmaceutics},
  109:93--102.

\bibitem[Perinelli et~al., 2020]{Perinelli2020}
Perinelli, D.~R., Cespi, M., Lorusso, N., Palmieri, G.~F., Bonacucina, G., and
  Blasi, P. (2020).
\newblock Surfactant self-assembling and critical micelle concentration: One
  approach fits all?
\newblock {\em Langmuir : the ACS journal of surfaces and colloids},
  36:5745--5753.

\bibitem[Perumal et~al., 2022]{Perumal2022}
Perumal, S., Atchudan, R., and Lee, W. (2022).
\newblock A review of polymeric micelles and their applications.
\newblock {\em Polymers}, 14.

\bibitem[Qin et~al., 2021]{Qin2021}
Qin, S., Jin, T., Van~Lehn, R.~C., and Zavala, V.~M. (2021).
\newblock Predicting critical micelle concentrations for surfactants using
  graph convolutional neural networks.
\newblock {\em The Journal of Physical Chemistry. B}, 125:10610--10620.

\bibitem[Rittig et~al., 2023a]{Rittig2022}
Rittig, J.~G., {Ben Hicham}, K., Schweidtmann, A.~M., Dahmen, M., and Mitsos,
  A. (2023a).
\newblock Graph neural networks for temperature-dependent activity coefficient
  prediction of solutes in ionic liquids.
\newblock {\em Computers and Chemical Engineering}, 171:108153.

\bibitem[Rittig et~al., 2023b]{Rittig2023_GDI}
Rittig, J.~G., Felton, K.~C., Lapkin, A.~A., and Mitsos, A. (2023b).
\newblock Gibbs-{D}uhem-informed neural networks for binary activity
  coefficient prediction.
\newblock {\em Digital Discovery}, 2:1752--1767.

\bibitem[Rittig et~al., 2022]{rittig2022graph}
Rittig, J.~G., Gao, Q., Dahmen, M., Mitsos, A., and Schweidtmann, A.~M. (2022).
\newblock Graph neural networks for the prediction of molecular
  structure-property relationships.
\newblock arXiv preprint arXiv:2208.04852.

\bibitem[Rosen and Kunjappu, 2012]{Rosen2012}
Rosen, M. and Kunjappu, J. (2012).
\newblock {\em Surfactants and Interfacial Phenomena: Rosen/Surfactants 4E}.
\newblock John Wiley \& Sons, Ltd.

\bibitem[Rosen et~al., 1982a]{Rosen1982a}
Rosen, M.~J., Cohen, A.~W., Dahanayake, M., and Hua, X.~Y. (1982a).
\newblock Relationship of structure to properties in surfactants. 10. surface
  and thermodynamic properties of 2-dodecyloxypoly(ethenoxyethanol)s,
  c12h25(oc2h4)xoh, in aqueous solution.
\newblock {\em J. Phys. Chem.}, 86(4):541--545.

\bibitem[Rosen et~al., 1982b]{Rosen1982}
Rosen, M.~J., Dahanayake, M., and Cohen, A.~W. (1982b).
\newblock Relationship of structure to properties in surfactants. 11. surface
  and thermodynamic properties of n-dodecyl-pyridinium bromide and chloride.
\newblock {\em Colloids and Surfaces}, 5(2):159--172.

\bibitem[Ruder, 2017]{ruder2017overview}
Ruder, S. (2017).
\newblock An overview of multi-task learning in deep neural networks.
\newblock arXiv preprint arXiv:1706.05098.

\bibitem[Sanchez~Medina et~al., 2022]{SanchezMedina2022}
Sanchez~Medina, E.~I., Linke, S., Stoll, M., and Sundmacher, K. (2022).
\newblock Graph neural networks for the prediction of infinite dilution
  activity coefficients.
\newblock {\em Digital Discovery}, 1:216--225.

\bibitem[Schramm et~al., 2003]{Schramm2003}
Schramm, L., Stasiuk, E., and Marangoni, G. (2003).
\newblock Surfactants and their applications.
\newblock {\em Annu. Rep. Prog. Chem., Sect. C: Phys. Chem.}, 99:3--48.

\bibitem[Schweidtmann et~al., 2020]{Schweidtmann2020}
Schweidtmann, A.~M., Rittig, J.~G., König, A., Grohe, M., Mitsos, A., and
  Dahmen, M. (2020).
\newblock Graph neural networks for prediction of fuel ignition quality.
\newblock {\em Energy \& Fuels}, 34(9):11395--11407.

\bibitem[Seddon et~al., 2022]{Seddon2022}
Seddon, D., Müller, E.~A., and Cabral, J.~T. (2022).
\newblock Machine learning hybrid approach for the prediction of surface
  tension profiles of hydrocarbon surfactants in aqueous solution.
\newblock {\em Journal of Colloid and Interface Science}, 625:328--339.

\bibitem[Shaban et~al., 2020]{Shaban2020}
Shaban, S.~M., Kang, J., and Kim, D.-H. (2020).
\newblock Surfactants: Recent advances and their applications.
\newblock {\em Composites Communications}, 22:100537.

\bibitem[Simonovsky and Komodakis, 2017]{Simonovsky2017}
Simonovsky, M. and Komodakis, N. (2017).
\newblock Dynamic edge-conditioned filters in convolutional neural networks on
  graphs.
\newblock In {\em 2017 IEEE Conference on Computer Vision and Pattern
  Recognition (CVPR)}, pages 29--38.

\bibitem[Su et~al., 2020]{Su2020}
Su, H., Wang, F., Ran, W., Zhang, W., Dai, W., Wang, H., Anderson, C.~F., Wang,
  Z., Zheng, C., Zhang, P., Li, Y., and Cui, H. (2020).
\newblock The role of critical micellization concentration in efficacy and
  toxicity of supramolecular polymers.
\newblock {\em Proceedings of the National Academy of Sciences of the United
  States of America}, 117:4518--4526.

\bibitem[Szűts and Szabó-Révész, 2012]{Szuets2012}
Szűts, A. and Szabó-Révész, P. (2012).
\newblock Sucrose esters as natural surfactants in drug delivery systems—a
  mini-review.
\newblock {\em International Journal of Pharmaceutics}, 433(1):1--9.

\bibitem[Tadros, 2005]{2005Surf}
Tadros, T. (2005).
\newblock {\em Surfactants in Personal Care and Cosmetics}, chapter~12, pages
  399--432.
\newblock John Wiley \& Sons, Ltd.

\bibitem[Thiruvengadam et~al., 2020]{Thiruvengadam2020}
Thiruvengadam, S., Murphy, M., Tan, J.~S., and Miller, K. (2020).
\newblock A generalized theoretical model for the relationship between critical
  micelle concentrations, pressure, and temperature for surfactants.
\newblock {\em Journal of Surfactants and Detergents}, 23(2):273--303.

\bibitem[Thompson et~al., 2023]{Thompson2023}
Thompson, C.~J., Ainger, N., Starck, P., Mykhaylyk, O.~O., and Ryan, A.~J.
  (2023).
\newblock Shampoo science: A review of the physiochemical processes behind the
  function of a shampoo.
\newblock {\em Macromolecular Chemistry and Physics}, 224(3):2200420.

\bibitem[Vieira et~al., 2021]{Vieira2021}
Vieira, I. M.~M., Santos, B. L.~P., Ruzene, D.~S., and Silva, D.~P. (2021).
\newblock An overview of current research and developments in biosurfactants.
\newblock {\em Journal of Industrial and Engineering Chemistry}, 100:1--18.

\bibitem[Wang et~al., 2019]{Wang2019}
Wang, Z., Chen, Y., Zhang, F., and Lin, S. (2019).
\newblock Significance of surface excess concentration in the kinetics of
  surfactant-induced pore wetting in membrane distillation.
\newblock {\em Desalination}, 450:46--53.

\bibitem[Weininger, 1988]{Weininger1988}
Weininger, D. (1988).
\newblock Smiles, a chemical language and information system. 1. introduction
  to methodology and encoding rules.
\newblock {\em J. Chem. Inf. Comput. Sci.}, 28(1):31--36.

\bibitem[Wieder et~al., 2020]{Wieder2020}
Wieder, O., Kohlbacher, S., Kuenemann, M., Garon, A., Ducrot, P., Seidel, T.,
  and Langer, T. (2020).
\newblock A compact review of molecular property prediction with graph neural
  networks.
\newblock {\em Drug Discovery Today: Technologies}, 37:1--12.

\bibitem[Yang et~al., 2019]{Yang2019}
Yang, K., Swanson, K., Jin, W., Coley, C., Eiden, P., Gao, H., Guzman-Perez,
  A., Hopper, T., Kelley, B., Mathea, M., Palmer, A., Settels, V., Jaakkola,
  T., Jensen, K., and Barzilay, R. (2019).
\newblock Analyzing learned molecular representations for property prediction.
\newblock {\em Journal of Chemical Information and Modeling}, 59(8):3370--3388.
\newblock PMID: 31361484.

\bibitem[Yang and Brouillette, 2016]{Yang2016}
Yang, Z. and Brouillette, C.~G. (2016).
\newblock Chapter thirteen - a guide to differential scanning calorimetry of
  membrane and soluble proteins in detergents.
\newblock In Feig, A.~L., editor, {\em Calorimetry}, volume 567 of {\em Methods
  in Enzymology}, pages 319--358. Academic Press.

\bibitem[Zhang et~al., 2019]{Zhang2019}
Zhang, F., Li, S., Zhang, Q., Liu, J., Zeng, S., Liu, M., and Sun, D. (2019).
\newblock Adsorption of different types of surfactants on graphene oxide.
\newblock {\em Journal of Molecular Liquids}, 276:338--346.

\bibitem[Zhang and Yang, 2017]{Zhang2017}
Zhang, Y. and Yang, Q. (2017).
\newblock A survey on multi-task learning.
\newblock {\em IEEE Transactions on Knowledge and Data Engineering},
  34:5586--5609.

\bibitem[Zhuang et~al., 2020]{Zhuang2020}
Zhuang, F., Qi, Z., Duan, K., Xi, D., Zhu, Y., Zhu, H., Xiong, H., and He, Q.
  (2020).
\newblock A comprehensive survey on transfer learning.
\newblock {\em Proceedings of the IEEE}, PP:1--34.

\end{thebibliography}

\end{document}